\documentclass{eas}
\usepackage{graphicx}
\begin{document}

\title{Wide Field Astronomy at Dome C: two IR surveys complementary to SNAP} 
\author{Denis Burgarella}\address{OAMP/LAM, traverse du siphon, B.P. 8 13376 Marseille Cedex 12, France}
\author{Marc Ferrari}\address{OAMP/LAM, 2 place Le Verrier, Marseille Cedex 4, France}
\author{Thierry Fusco}\address{ONERA, BP 52, 29 avenue de la Division Leclerc, 92320 Chatillon Cedex, France }
\author{Maud Langlois}\address{OAMP/LAM, 2 place Le Verrier, Marseille Cedex 4, France}
\author{Gerard Lemaitre}\address{OAMP/LAM, 2 place Le Verrier, Marseille Cedex 4, France}
\author{Brice Le Roux}\address{OAMP/LAM, 2 place Le Verrier, Marseille Cedex 4, France}
\author{Gil Moretto}\address{OAMP/LAM, traverse du siphon, B.P. 8 13376 Marseille Cedex 12, France}
\author{Magalie Nicole}\address{ONERA, BP 52, 29 avenue de la Division Leclerc, 92320 Chatillon Cedex, France }
\begin{abstract}
Surveys provide a wealth of data to the astronomical community that are used well after their completion. In this paper, we propose a project that would take the maximum benefit of Dome C in Antarctica by performing two surveys, in the wavelength range from 1-5 $\mu m$, complementary to SNAP space surveys. The first one over 1000 sq. deg. (1 KdF) for 4 years and the second one over 15 sq. deg (SNAP-IR) for the next 4 years at the same time as SNAP 0.35-1.7 $\mu m$ survey. By using a Ground-Layer Adaptive Optics system, we would be able to recover, at the ice level and over at least half a degree in radius, the $\sim 300$ mas angular resolution available above the 30-m high turbulent layer. Such a survey, combining a high angular resolution with high sensitivities in the NIR  and MIR, should also play the role of a pre-survey for JWST and ALMA.
\end{abstract}
\maketitle

\section{What are the Characteristics of Dome C ?}

Lawrence et al. (2004) published new atmospheric coherence times, isoplanetic angles and seeings mesured at Dome C, Antarctica from 23 March to 05 May 2004. The natural seeing ($\epsilon$) is excellent with a mean value of $\epsilon \approx 0.3$ arcsec, i.e. much better than other astronomical sites on Earth. The isoplanetic angle ($\theta$) is also very good with a mean value of $\theta \approx 6$ arcsecs. Finally the coherence time ($\tau$) is slightly better than other ground-based sites: $\tau \approx 8$ milliseconds. Such exceptional values would permit carrying out very high quality observations on very wide fields of view.

Later studies but Agabi et al. (2006), in the first half of 2005, however, showed that if the seeing is excellent at an altitude $h \ge 30$m with $\epsilon = 0.36 \pm 0.19$ arcsec, it is modest with $\epsilon = 1.4 \pm 0.6$ arcsec at the ground level (e.g. $h \ge 8.5$m above the ice). However, the isoplanetic angle does not seem to change with the altitude $\theta = 4.7 \pm 2.6$ arcsec. The coherence time changes from at $8.6 \pm 7.1$ milliseconds at $h \ge 30$m to $8.6 \pm 7.1$ milliseconds at $h \ge 8.5$m.

Kenyon \& Storey (2006)  estimated the area of the sky observable from Dome C and compared it to the area of the sky observable from equatorial sites such as Mauna Kea. While about two thirds of the sky can be observed from the equator or Mauna Kea, only about $25 \%$ of the sky can be from Dome C whose coordinates are $75^o06'25''$ South and  $123^o20'44''$ East. However, $25 \%$ of the sky means about 10000 square degrees. Do we need more to carry out a wide-field survey ? 

Kenyon \& Storey (2006) also showed that the range of zodiacal contributions to the sky brightness in the V-band ($SB_V$) is, in average, better at Dome C that at Mauna Kea: at zenith $SB_V$ is fainter than 23.2 Vmag.arcsec$^{-2}$ at Dome C and $SB_V$ is brighter than < 23.2 Vmag.arcsec$^{-2}$ at Mauna Kea.

Finally, what is certainly the best known characteristic of Antarctica is the low ambiant temperature which is about$ -50 ^o C$ in winter and $- 25 ^o C$ in summer (http://www.ifremer.fr/ifrtp/pages/concor.html). These temperatures, combined with the low zodiacal light, mean deeper sensitivities.

In summary, Dome C appears as an exceptional site for astronomical observations at an altitude above $\sim 30m$: the natural seeing is excellent ($\epsilon \approx 0.3$) and almost comparable to space observatories even though the image quality strongly degrades at the ice level. Isoplanetic angles are larger by a factor of two to three at any altitude compared to other ground-based sites and reaches $\theta \approx 6$ arcsec. Low contributions from the zodiacal light and from the telescope itself will permit high angular resolution and deep observations, especially in the near-infrared (NIR) and mid-infrared (MIR) over quite a wide field of view if we can solve the issue related to the turbulence layer.

\section{A deep wide-field NIR+MIR high angular resolution survey}

Dome C qualifies as an exceptional site to carry out deep observations in the NIR+MIR wavelength range. Would it be possible to add a wide-field qualification and make Dome C  the best site on Earth to carry out a wide-field NIR+MIR high angular resolution survey ?

\subsection{A technological challenge}

Beyond the general idea presented above, we need to be more practical and study whether any technology exists that would allow to reach the 300 milliarcsec (mas) angular resolution in the NIR+MIR wavelength range over a wide field of view (i.e.  larger than 30 arcmin in radius) in Antarctica.

To get rid of the effects of the turbulent ground-layer, an obvious solution would be put the telescope at an altitude $> 30$m by building a tower, on which the telescope would stand. By doing so, astronomers should get back the desired $\sim 0.3$ arcsec seeing without any constraints on the observable field of view. The main drawback to this solution would be that we cannot control the image quality. One only can wait for the best conditions to occur. In the present project, we would like to make a full use of Dome C characteristics and obtain observational conditions close to space ones.

As already shown in the previous section, isoplanetic angles of $\theta \approx 6$ arcsecs are observed at Dome C in the visible (Aristidi et al. 2005). For a comparison, the best ground-based sites in Aristidi et al.'s list are Cerro Pachon at $\theta \approx 3$ arcsecs and Paranal's $\theta \approx 2$ arcsecs. This difference means that the probability to find natural guide stars increases by a factor of 10 as compared to these other sites. Moreover, the absolute value should increase with the wavelength.

\subsection{Ground-Layer Adaptive Optics}

Classic adaptive optics enables large telescopes to provide diffraction limited images, but their corrected field is restrained by the angular decorrelation of the turbulent wave-fronts. However many scientific goals would benefit from a wide and uniformly corrected field, even with a partial correction. Ground Layer Adaptive Optics (GLAO) systems are supposed to provide such a correction by compensating the lower part of the atmosphere only. Indeed, this layer is at the same time highly turbulent and isoplanatic on a rather wide field. A GLAO correction therefore appears as a good and adapted system in the specific case of Dome C. A more detailed description of GLAO can be found in Andersen et al. (2006).

To check whether GLAO could be a solution to reach our specifications, we performe two independant simulations based on two sets of assumptions. The first set (set\#1 hereafter) assumes a 4m telescope with a 8x8-actuators deformable mirror from the V-band to the K-band and a turbulence seeing on the ice of 1.1 arcsec. The second set (set\#2 hereafter) assumes a 2m telescope with a 10x10-actuators deformable mirror from the I-band to the K-band and a turbulence seeing of 1.9 arcsec that are carried out using PAOLA (Jolissaint, Veran, Conan 2006). Set\#1 can be seen as a relatively optimistic simulation while set\#2 would be more pessimistic. The number of actuators is a very reasonable number (see Rabaud et al.'s paper in this same volume). These simulations seem encouraging. The gain in encircled energy is still larger than 6 at a radius of 30 arcmin in V, R, J, H and K (Fig. 1a) which is confirmed in Fig. 1b in I, J, H and K. If we trust these simulations, that means that GLAO is extremely useful in terms of encircled energy for a site like Dome C. 

Fig. 2 shows that the full width half maximum (FWHM) is lower than 300 mas and stable within a radius of 30 arcmin, in the NIR  i.e. within our specifications. Figs. 3a show how the Point Spread Function (PSF) would vary inside  fields of view  increasing in size as a function of the wavelength for set\#1. Note that this is not the PSF at the edge of the field but the average PSF over the considered field of view. On the other hand, Fig. 3b presents the PSF at different radii. Those results suggest that we can increase the size of the corrected GLAO field without degrading the PSF which is relatively constant within a given field.

\begin{figure}
 \includegraphics[width=6cm] {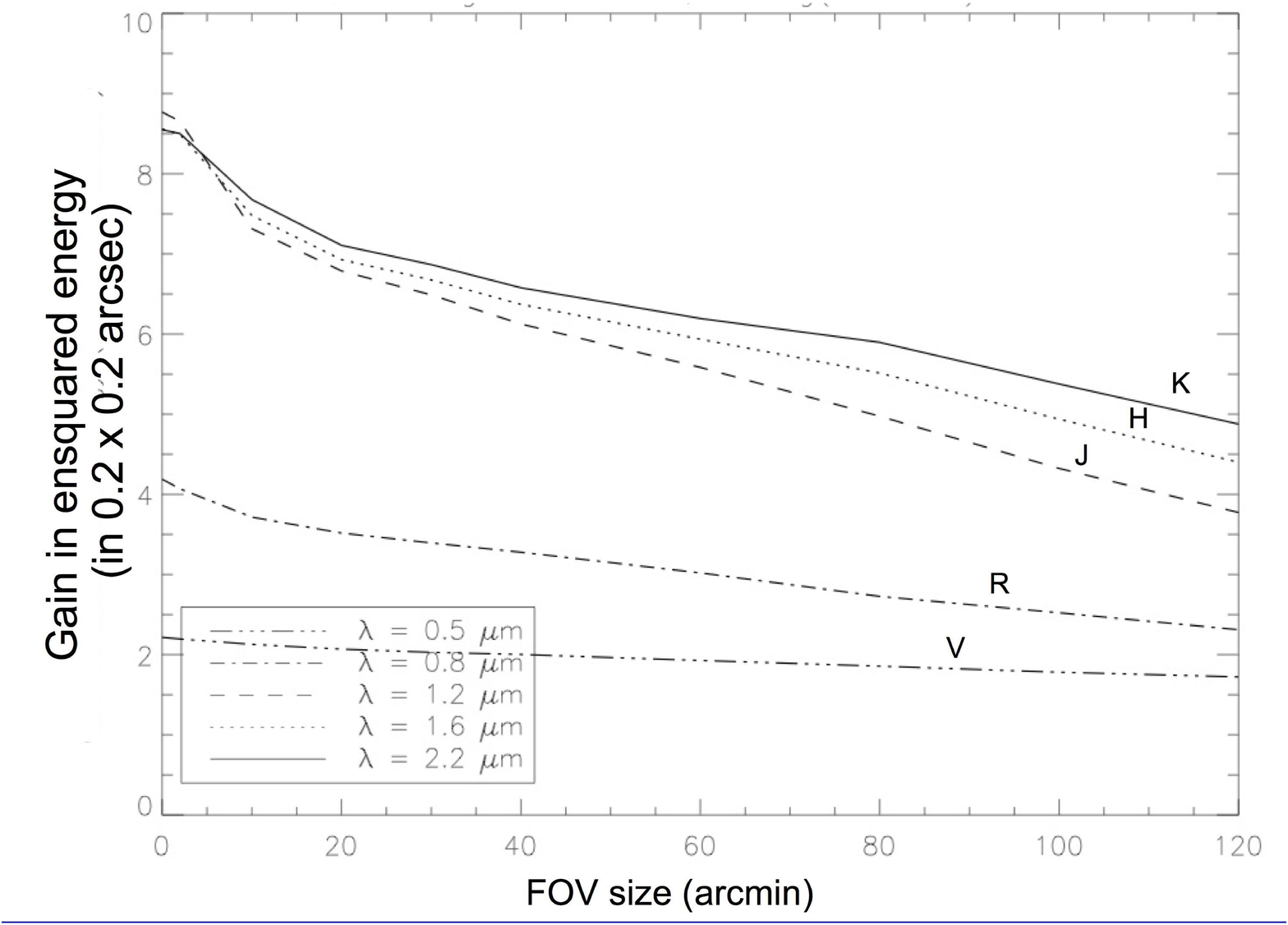}
\qquad
 \includegraphics[width=6cm, angle=0] {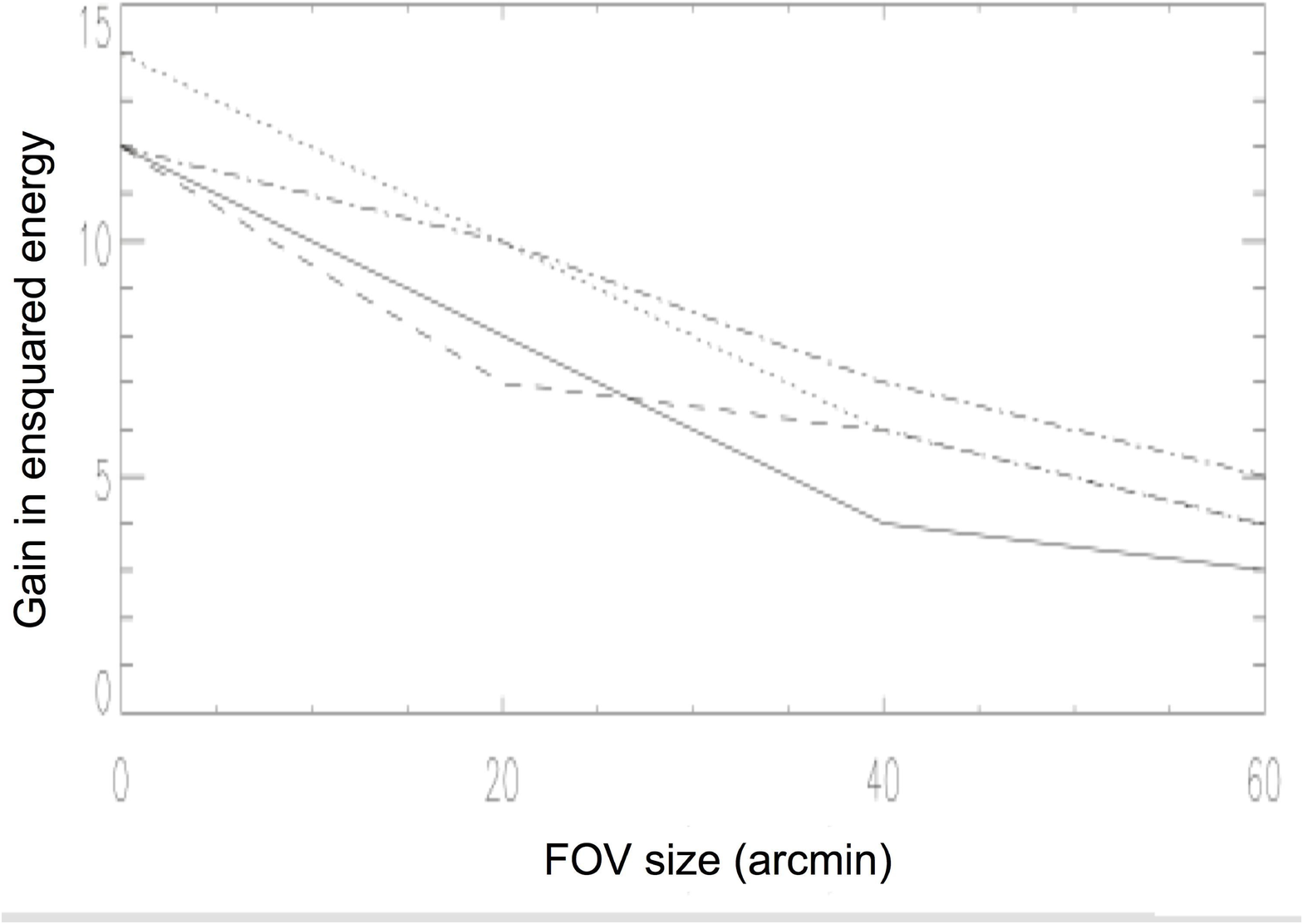}
\caption{the gain in encircled energy within a 0.2 x 0.2 arcsec$^2$ box in V, R, J, H \& K from GLAO simulation set\#1 (a) and set\#2 (b) is always larger than 6 up to a radius of 30 arcmin.}
\end{figure}

\begin{figure}
\center
 \includegraphics[width=6cm, angle=0] {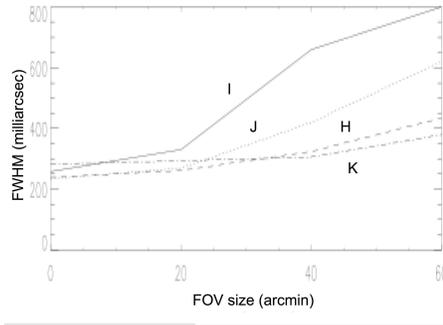}
\qquad
\caption{Simulations set\#2 suggest that the FWHM is stable and within our specifications for the three NIR bands if we assume a field of view with a maximum radius of 30 arcmin.}
\end{figure}

\begin{figure}
\center
 \includegraphics[width=8cm, angle=0] {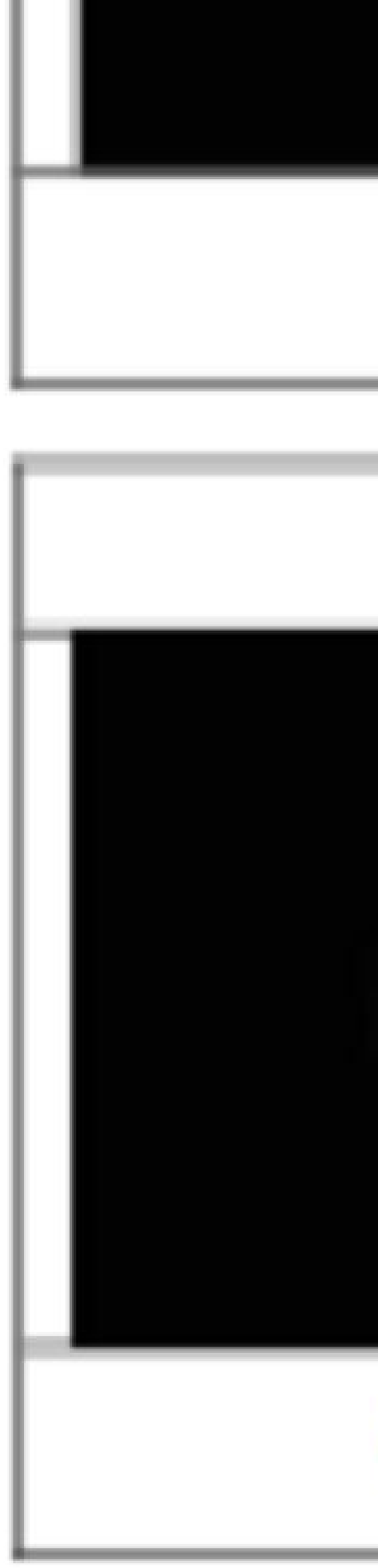}
\qquad
 \includegraphics[width=7cm, angle=0] {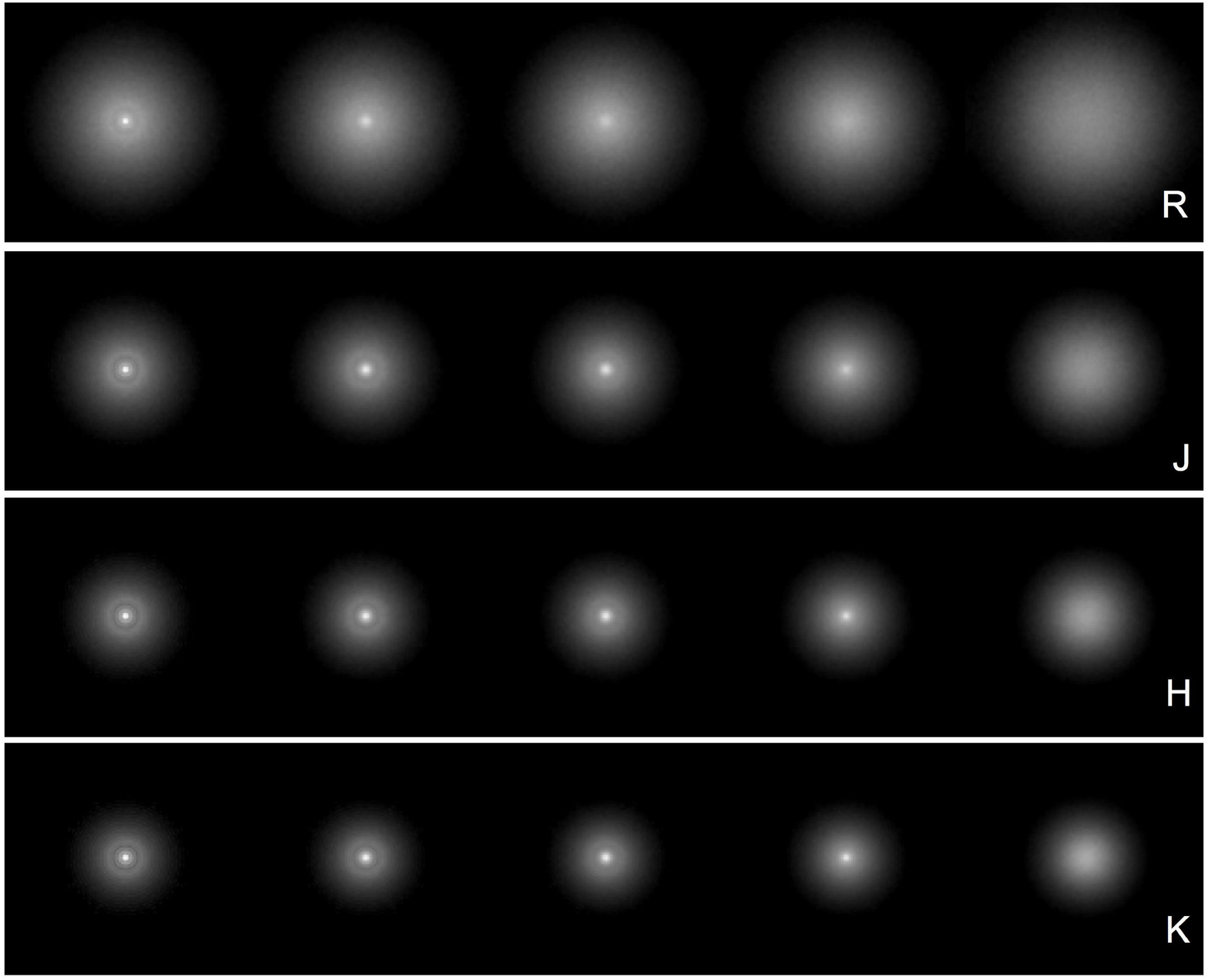}
\caption{a) For simulation set\#1, the average PSF within fields of view of 10', 30', 60' and 120' from the shortest to the longest wavelength (R to K) from top to bottom. The last column on the right hand side of the figure represent the natural seeing set to 1.1 arcsec. b) For simulation set\#2, the variation of the PSF from the core of the field to the outskirts from left to right and from the shortest to the longest wavelength from top to bottom. The last column on the right hand side of the figure represent the natural seing set to 1.9 arcsec.}
\end{figure}

\subsection{An additional gain: the OH suppressor}

The sky brightness in the wavelength range $0.65< \lambda (\mu m)< 0.22$ is dominated by OH emission. This OH emission  will be essentially identical at Dome C compared to that at all other sites on Earth. Decreasing the airglow would mean an increase of the signal-to-noise ratio by a factor of 2.5 in J and H or to increasing the limiting magnitude by $> 1$mag  (Iwamuro et al. 2001). Moreover, by decreasing the background due to the airglow, we can increase the maximum exposure times before saturation, which translates into a better efficiency of the telescope to build the survey.

\subsection{The telescope design}

Two types of telescopes are under study at the Observatoire Astronomique Marseille Provence that would allow to observe in the NIR+MIR over a wide field of view while not degrading the image quality above 0.3 arcsec. A first one is the Three-Reflection Telescope (TRT) designed by Lemaitre et al. ( 2004). The size of the spot is still 0.25 arcsec RMS at a radius of 1 degree, which is well within our specifications. Two down-scaled models of these telescopes already exist at IAS, Frascati and at OAMP/LAM, Marseille. Another option would be an off-axis design adapted from space telescope designs (e.g. Moretto et al. 2004) which bears several advantages: no obstruction of the beam by the secondary mirror, no diffraction / emission from the secondary support structure to degrade images / sensitivities.

\subsection{What are the expected performances of such a telescope ?}

We assume a 2.5m telescope and Burton et al. (2005) background in the NIR+MIR. The temperature of the telescope would be 200K with a $3 \%$ emissivity. Given that we would like a 0.3 arcsec angular resolution, we assume a 0.15-arcsec pixel scale. The telescope thoughput is set to $30 \%$  and finally, GLAO and an OH suppressor would be installed on the telescope. Fig. 4 gives the assumed angular resolution (the telescope would be diffraction limited above about $3 \mu m$), limiting magnitudes (SNR = 5 in 0.5h) for the very wide survey. If we assume a 1 sq. deg. survey (i.e. 1KdF). The numbers are in the same range that those published by Burton et al. (2005). For the deep survey (SNAP-IR), we assume a 15 sq. deg field of view (about 3.9 x 3.9 sq. deg) with SNR = 5. Since we wish to reach much deeper limiting magnitudes, exposure times are set to 26h, each instantaneous field would be observed with a 4-day frequency.

\begin{figure}
\center
 \includegraphics[width=8cm, angle=0] {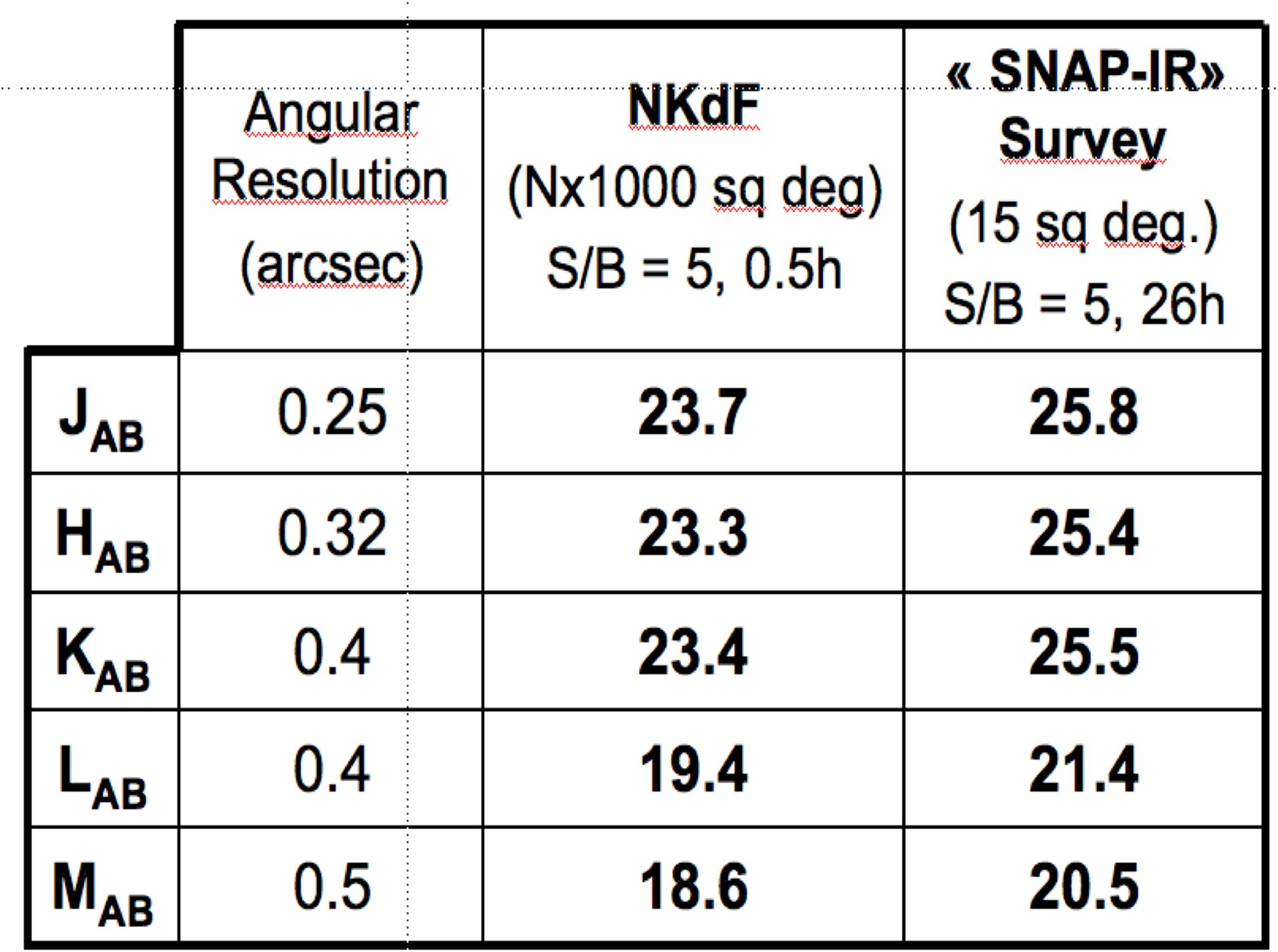}
\qquad
\caption{Estimated performances for the two surveys 1KdF and SNAP-IR for 1 to 5 $\mu m$.}
\end{figure}

The instantaneous field of view would be 0.7 x 0.7 sq. deg (0.5 sq. deg but money-dependent) which correspond to 30 individual fields. If we want to complete 1KdF in 4 years with a 0.5h exposure time per individual field , we need (75 \% observational efficiency i.e. 1325 hours per year: Sadibekova et al. 2005) 1325 x 4 = 5300 hours to complete the deep survey.

\section{Quick tour of the Science Case for such a survey}

We will have a SDSS/VISTA-like field of view and a SNAP-like angular resolution from 1 to 5 $\mu m$. Quite a number of science topics can take advantage of those figures. Given the limited size of this paper, we will only highlight a few of them. Of course, beyond the pure science, such surveys would be very helpful as pre-surveys for JWST and ALMA.

The 1KdF and SNAP-IR surveys will very likely have a direct impact on the detection and the analysis of very large samples of high redshift galaxies in the NIR + MIR (e.g. Aldering et al. 2005): 

- from resolved galaxies observed in the rest-frame optical, we will be able to study the mass assembly and morphogenesis of galaxies (e.g. Bundy et al. 2006),

- cosmic shear measurements will benefit from the good and stable PSF over wide fields of view (Mellier et al. 2002),

- supernovae detections will be possible in dusty environments where most of the star formation happens at $1 < z < 2$ (Riess et al., 2004; Elbaz et al. 2002),

- a mapping of the Magellanic clouds in the NIR/MIR from the 1KdF survey would permit to study star formation regions in our two satellites,

- taking advantage of the 0.5-h individual fields over 4 years, it might also be possible (although cosmological fields are selected away from the Milky Way disk) to look for hot-Jupiters and hot-Neptunes around M stars in the SNAP-IR survey.

\section{A complement to SNAP in the NIR-MIR ?}

SNAP (e.g. Aldering 2005) will carry out two multi-band surveys down to $m_{AB} = 30.3$ over 15 sq. deg and $m_{AB} = 27.7$ over 300-1000 sq. deg in the wavelength range 0.35 - 1.7 $\mu m$. The telescope would feature a 2-m primary mirror and a pixel scale of 0.1 arcsec. Our two surveys would bring complementary data to SNAP:

- data beyond $\lambda =1.7 \mu m$, would be very useful to estimate photometric redshifts (e.g. Rowan-Robinson et al. 2005).

- the high sensitivity in the NIR/MIR opens up a new population of dusty supernovae that can hardly be observed in the visible range (e.g. Maiolino et al. 2002). SNAP will detect supernovae up to a redshift of $z \sim 1 - 2$. We know that most of the star formation in the universe in this redshift range is hidden in dust (e.g. Takeuchi, Buat, Burgarella 2004). The detection of supernovae in dusty environment is therefore a very promising way to explore and they have been more and more studied recently (e.g. Chary et al. 2005, Pozzo et al. 2006, Elias-Rosa et al. 2006).

\section{Conclusions}

Dome C provides an excellent opportunity to perform a very wide field imaging survey in the wavelength range $1 - 5 \mu m$. 

The use of Ground Layer Adaptive Optics will allow to control the image quality and reach a FWHM of 300 milliarcsecs up to a 1 sq. deg. diameter field of view.

Thanks to the intrinsic low background (average temperature of $-50^o C$ and low zodiacal light) and the use of an OH suppressor device, we would be able to carry two surveys. A first one would cover at least 1000 sq. deg. (1KdF) down to $m_{AB} \sim 23 - 24$ in JHK and $m_{AB} \sim 19$ above 3 $\mu m$. A second one would cover 15 sq. deg. (SNAP-IR) down to $m_{AB} \sim 25 - 26$ in JHK and $m_{AB} \sim 21$ above 3 $\mu m$. This would provide a very nice complement to SNAP and should play the role of a pre-survey for JWST and ALMA.

This project would provide two low-cost surveys (compared to space standards). Given the environment, any facility in Antarctica would benefit from simple operations which match the survey modes presented in this paper.



\begin{thebibliography}{99}
\bibitem[2006]{Agabi} Agabi A., Aristidi E., Azouit M., Fossat E., Martin F. et al. 2006, PASP 118, 344
\bibitem[2005]{Aldering} Aldering G. 2005, New AR 49, 346
\bibitem[2006]{Andersen} Andersen D.R., Stoesz J., Morris S. Lloyd-Hart, M., Crampton T. et al. 2006, astro-ph/0610097
\bibitem[2006]{Aristidi} Aristidi E., Agabi A., Fossat E., Azouit M., Martin F. et al. 2005, A\& A 444, 651 
\bibitem[2005]{Bundy} Bundy K, Ellis R.S., Conselice C.J., Taylor J.E., Cooper M.C. et al. 2006, ApJ 651, 120
\bibitem[2005]{Burton} Burton M.G., Lawrence J.S., Ashley M.C.B., Bailey J.A, Blake C. et al. 2005, PASA 22, 199
\bibitem[2005]{Chary} Chary R., Dickinson M.E., Teplitz H.I., Pope A., Ravindranath S. 2005, ApJS 635, 1022
\bibitem[2005]{Elbaz} Elbaz D., Cesarsky C.J., Chanial P., Aussel H., Franceschini A. 2002, A\& A 384, 848
\bibitem[2005]{Elias-Rosa} Elias-Rosa N., Benetti S., Cappelaro E., Turatto M., Mazzali P.A. 2006, MNRAS 369, 1880
\bibitem[2006]{Iwamuro} Iwamuro F., Motohara K., Maihara T., Iwai J., Tanabe H. et al. 2001, PASJ 53, 355 
\bibitem[2006]{Jolissaint} Jolissaint L., V\'eran J.-P., Conan R. 2006, JOSA A 23, 382
\bibitem[2006]{Kenyon} Kenyon S.L. \& Storey J.W.V. 2006, PASP 118, 489 
\bibitem[2004]{Lawrence} Lawrence J.S., Ashley M.C.B., Tokovonin A., Travouillon T. 2004, Nature 431, 278 
\bibitem[2006]{Lemaitre} Lemaitre G.R., Montiel, P., Joulie, P., Dohlen, K., Lanzoni, P. 2004, SPIE 5494, 426  
\bibitem[2002]{Maiolino} Maiolino R., Vanzi L., Mannucci F., Cresci G., Ghinassi F., Della Valle M. 2002, A\& A 389, 84
\bibitem[2002]{Mellier} Mellier Y., van Waerbeke L., Bertin E., Tereno I., Bernardeau F. 2002, SPIE 4847, 112
\bibitem[2006]{Moretto} Moretto G., Langlois, M., Ferrari, M. 2004, SPIE 5487, 1111
\bibitem[2006]{Pozzo} Pozzo M., Meikle W.P.S., Rayner J.T., Joseph R.D., Filippenko A.V. et al. 2006, MNRAS 368, 1169
\bibitem[2006]{Rabaud} Rabaud D. et al. 2006, this volume
\bibitem[2006]{Riess} Riess A.G., Strolger L.-G., Tonry J., Casertano S., Ferguson H.C. et al. 2004, ApJ 607, 665
\bibitem[2004]{Rowan-Robinson} Rowan-Robinson M., Babbedge T., Surace J., Shupe D. Fang F. et al. 2005, AJ 129, 1183
\bibitem[2005]{Sadibekova} Sadibekova T., Agabi E., Aristidi E., Azouit M. Chadid M. et al. 2006, this volume
\bibitem[2004]{Takeuchi} Takeuchi T.T., Buat V., Burgarella D. 2004, A \& A 440, L17
\end{thebibliography}
\end{document}